\documentstyle[aps,preprint,eqsecnum]{revtex}
\begin{document}
\draft
\preprint
{$\hbox to 5 truecm{\hfill CGPG-96/9-5}\atop
{\hbox to 5 truecm{\hfill gr-qc/9609048}\atop}$}
\title{The postulates of gravitational thermodynamics}
\author{Erik A. Martinez 
\footnote{electronic address:
martinez@phys.psu.edu}}
\address{Center for Gravitational Physics and Geometry,\\
Department of Physics,\\
The Pennsylvania State University,\\
University Park, PA 16802-6300, USA}
\maketitle
\begin{abstract}
\noindent{\baselineskip=14pt
The general principles and logical structure of a thermodynamic 
formalism that incorporates strongly self-gravitating systems are
presented. This framework generalizes and simplifies the formulation of 
thermodynamics  developed by Callen.  The definition of
extensive variables, the homogeneity properties of intensive
parameters, and  the fundamental problem of gravitational thermodynamics
are discussed in detail. In particular, extensive parameters include
quasilocal quantities and are naturally incorporated into a set of basic
general postulates for thermodynamics.  These include additivity of
entropies (Massieu functions) and the generalized second law. Fundamental
equations are no longer homogeneous first-order functions of their
extensive variables. It is shown that the postulates lead to a formal
resolution of the fundamental problem despite non-additivity of extensive
parameters and thermodynamic potentials.  Therefore, all the results of
(gravitational) thermodynamics are an outgrowth of these
postulates. The origin  and  nature of the differences with ordinary
thermodynamics are analyzed. Consequences of the formalism include the
(spatially) inhomogeneous character of thermodynamic equilibrium states,
a reformulation of the Euler equation, and the absence of a Gibbs-Duhem
relation.  
\par} 
\end{abstract} \vspace{7mm}
\pacs{PACS numbers: 04.70.Dy, 04.40.-b, 97.60.Lf}
\newpage

\section{Introduction}

The inclusion of gravity into a physical theory fundamentally
alters its basic assumptions and structure. 
Perhaps the most  general and universal of all theories is
thermodynamics. It is therefore particularly interesting to understand
the changes induced by gravity in its underlying principles and
to organize them into a general framework that serves as the foundation
of a thermodynamic theory that incorporates strongly self-gravitating
systems. Although this is relevant in its own right, it is also
important in a somewhat different context: gravitational
thermodynamics is expected to arise as one of the macroscopic limits of
quantum gravity.  Despite the fact that progress continues to be made
into the way gravity alters the structure of quantum field theories
\cite{AsLe}, there does not exist yet a complete theory of quantum
gravity.  We believe that the characteristic features and principles of a
theory of  gravitational thermodynamics have to be understood 
properly before they can be fully justified by means of statistical
methods derived from one or several candidate theories of quantum gravity.

It is our purpose in this paper to formulate the general
principles of gravitational thermodynamics.
We shall discuss in detail the basic definitions and concepts of this
formalism, its overall structure,  its fundamental problem, the minimal
set of assumptions (the postulates) that lead to the formal resolution of
this problem as well as their mathematical and physical consequences. 
This formulation  clarifies not only the  differences and similarities
between gravitational and  nongravitational thermodynamics
within a coherent framework, but also a  number of existing misconceptions
concerning its logical foundations and results.  
It generalizes the overall structure of thermodynamic theory and may
provide a  basic framework to incorporate current and future progress
\cite{AsLe,qg} in a statistical mechanical description  of
self-gravitating systems.

A clear and elegant formulation of the foundations and
structure of ordinary (that is, nongravitational)  thermodynamics has
been introduced by Callen \cite{Ca}.   However, this formalism cannot
describe the thermodynamics of gravitational systems in its present form
and needs to be modified.   To recognize this, we shall follow the
following strategy.  We believe that the most effective way to gain
insight into  the limitations of a set of physical principles 
is by studying a model system that illustrates them without unnecessary
complications. We introduce such a model problem in Sec. II.  It  is
the simplest example of a composite self-gravitating system at finite
temperature and  resembles as closely as possible the textbook example of
a composite system in ordinary thermodynamics.  The model  is
general enough to capture all the non-trivial behavior of gravity, but
simple enough to allow an exact evaluation of all quantities, and 
provides insight into the places where the thermodynamic formalism has to
be refined.  As a preparatory step in the study of the postulatory basis
of thermodynamics, we show that it provides
a counterexample to a basic assertion \cite{Ca,GrCa} of ordinary
thermodynamics: in gravitational thermodynamics, additivity of entropies 
applied to spatially separate subsystems does {\it not} depend on or
require the  entropies of the latter to be homogeneous first-order
functions of their extensive parameters.

The general principles of gravitational thermodynamics are developed
from the start in Sec. III, which constitutes the core of the
paper. It is shown that a general and rigorous characterization of the
defining properties of thermodynamical self-gravitating systems
originates from two factors, namely (1) the particular characteristics of
their extensive variables, and (2) the homogeneous properties of their
intensive parameters as functions of extensive variables. The correct
definitions of these variables are discussed in detail.
Extensive variables include quasilocally defined quantities like
quasilocal energy and provide the background for the postulates.
We discuss the composition of thermodynamic systems, the existence of
fundamental equations, and  formulate the fundamental problem of
gravitational thermodynamics. Our object is to follow 
as closely as possible the logic of Callen's formalism and  generalize
its basic definitions and  postulates wherever it is required.  
The definitions and postulates arrived at are a natural extension of 
the ones of ordinary thermodynamics with  appropriate modifications
necessary to accommodate the global aspects and nonlinearities
characteristic of gravity. The postulates are the basic principles of
thermodynamics when strongly self-gravitating systems are considered.  
We revise completely the logic of Sec. II and show that  additivity of
entropies (and, in general, additivity of all Massieu  functions) as well
as the generalized first and second laws of thermodynamics are part of
these postulates and, as such, are not to be proven within the
thermodynamic formalism.  We demonstrate that the fundamental problem is
formally solved by these  postulates in spite of  strong
interaction between  constituent subsystems and the associated
nonadditivity of extensive variables. The preeminence of Massieu
functions over thermodynamic potentials is discussed. The conditions of
thermodynamical equilibrium  and its  (spatially) inhomogeneous nature
(equivalence principle) are a direct result of the postulates.

Although the fundamental equation and the intensive variables maintain
their mutual relationships, the former is no longer a homogeneous
first-order function of  extensive parameters. 
This property is not a consequence of
the inhomogeneity of equilibrium configurations but of
the functional form of the  gravitational equations of state. These are
in general no longer homogeneous zeroth-order functions  under rescaling
of extensive parameters.  We show that these central
differences with nongravitational thermodynamics are not forbidden by
the logical structure of the formalism and can be incorporated easily
into it by relaxing several assumptions in the postulates.  
However, the mathematical consequences  of the new
postulates are quite different from the ones familiar in
regular thermodynamics. Formal relationships (like the Euler
equation) must be reformulated and there is no direct analogue  of the
ordinary  Gibbs-Duhem relation among intensive
variables. Nonetheless, this does not influence significantly the
applicability of the thermodynamic formalism.

The principles of the framework and its logic apply to general 
systems at finite temperature. 
They incorporate the so-called gravothermal catastrophe and other
peculiar thermodynamical behavior observed in self-gravitating objects. 
As such, {\it all} standard equilibrium thermodynamics of gravitational
configurations is a consequence of the general postulates presented here. 
The approach generalizes ordinary thermodynamics and provides  a
consistent treatment of composite self-gravitating systems (with or
without matter fields). Moreover, it  becomes evident that the
modifications in the thermodynamic formalism necessary to incorporate 
gravity  liberates it from assumptions  that appeared (and are not)
fundamental, and as such, allows us to regard its general character and
the extent of its logic in their full measure.

\section{Homogeneity and additivity} 

We motivate our presentation of the postulates  of
gravitational thermodynamics and their consequences with a model
problem.  It consists of a spherically symmetric uncharged black hole
surrounded  by matter (represented by a thin shell).  
We  shall find in this section the
relationship among the entropy functions $S_0$, $S_B$, and $S_S$ whenever
thermodynamic equilibrium is achieved and its connection with the
functional dependence of those functions. 
In what follows, subscripts $B$ and $S$ refer to quantities for  black
hole and shell constituents respectively, whereas the subscript $0$ refers
to quantities for the  total system.

The composite system is characterized by the surface area 
$A_0 = 4\pi r_0^2$ of a  two-dimensional spherical  boundary
surface $B_0$ (located at $r=r_0$) that encloses the components,  and
the quasilocal energy $E_0$  contained within \cite{MaYo,Yo}. 
The Arnowitt-Deser-Misner (ADM) mass $m_+$ of the system as a 
function of these variables is 
\begin{equation}
m_+ (E_0, A_0) = E_0 \, \Bigg( 1 - {{E_0}\over{2r_0}} \Bigg)\ . 
\label{m+}
\end{equation}
Throughout the paper, units are chosen so that
$G =\hbar = c = k_{Boltzmann} =1$. Due to spherical symmetry of the
model problem,  we use areas and radii interchangeably.
The pressure $p_0$ associated to the surface $B_0$ is obtained as 
(minus) the partial derivative of the energy $E_0$ with respect to $A_0$
at constant $m_+$. It corresponds to a negative surface tension and reads
\begin{equation}
	p_0 (E_0, A_0) = {{{E_0}^2}\over{16 \pi {r_0}^3}}
	\Bigg(1 - {{E_0}\over{r_0}} \Bigg)^{-1} = \,
	{{1}\over{16\pi \, r_0 \,  k_0}} (1 - k_0)^2 \ , \label{p0}
\end{equation}
where $k_0 \equiv (1 - 2m_+/r_0)^{1/2}$.  
Let $\beta_0$ denote the inverse temperature 
at the surface $B_0$. The first law of thermodynamics for the  system
reads 
\begin{equation}
	dS_0 = \beta_0 \, (dE_0 +  p_0 \, dA_0) \ ,   
\end{equation}
which can be written as a total differential  by using Eqs. (\ref{m+})
and (\ref{p0})  as
\begin{equation}
	dS_0 = \beta_0 \, {k_0}^{-1} \,  dm_+ \ . \label{ds0}
\end{equation}

Consider now the constituent systems. The black hole is characterized
by the surface area $A_B = 4\pi R^2$ of a spherical boundary
surface $B_R$ located at $r=R \leq r_0$, and by its 
quasilocal energy $E_B$. 
The  ADM mass $m_-$ of the black hole as a function of
these variables is  \cite{Yo}
\begin{equation}
m_- (E_B, A_B) = E_B \, \Bigg( 1 - {{E_B}\over{2R}} \Bigg) \ . \label{m-}
\end{equation}
The horizon radius of the black hole is $2m_-$. The 
different radii satisfy the inequalities $2m_- \leq 2m_+ \leq R \leq r_0$,
where $2m_+$ represents the horizon radius 
of the total system.  
The pressure associated to the gravitational field of the black hole at
the surface $B_R$ is
\begin{equation}
	p_B (E_B, A_B) = {{{E_B}^2}\over{16 \pi {R}^3}} 
	\Bigg(1 - {{E_B}\over{R}} \Bigg)^{-1} = \,
	{{1}\over{16 \pi R\, k_-}} (1 - k_-)^2 \ , \label{pb}
\end{equation}
where $k_- \equiv (1 - 2m_- /R)^{1/2}$.
As it is well known, 
the first law of thermodynamics for the black hole can be 
expressed as a total differential by using Eqs. (\ref{m-}) and 
(\ref{pb}), namely
\begin{eqnarray}
	dS_B &=& \beta_B \, (dE_B +  \, p_B \, dA_B)  \nonumber \\
	     &=& \beta_B \, {k_-}^{-1} \, dm_-  \ , \label{dsb}
\end{eqnarray}
where $\beta_B$ is the inverse temperature of the
black hole at the  surface $B_R$.  This equation must be contrasted
with the entropy differential  (\ref{ds0}) for the total system.

For thermodynamical purposes, the shell is considered (effectively) at
rest. For simplicity, we assume that the shell surface
coincides with the surface  $B_R$. (This does not
imply loss of generality \cite{MaYo}.)  
Thus, the surface areas for
the black hole and shell coincide: $A_B = A_S \equiv  A_R = 4\pi R^2$. 
The gravitational junction conditions \cite{Is,La} at the shell position 
determine its surface energy density $\sigma$ and surface pressure $p_S$
to be 
\begin{equation} 
	E_S \equiv 4\pi R^2 \sigma = R \, (k_- - k_+) \ , \label{m}
\end{equation}
and
\begin{equation}
	p_S = {{1}\over{16 \pi R \, k_- \,  k_+}} \, 
	\Big[ k_- (1-k_+)^2 - k_+  
	(1-k_-)^2 \Big] \ , \label{ps}
\end{equation}
respectively. The symbol $E_S$ denotes the local mass-energy of matter
and $k_+ \equiv (1 - 2m_+ /R)^{1/2}$. To  simplify the analysis further,
we consider only  the case when the total number of particles $N_S$ 
in the shell is constant.  The condition $m_+ \geq m_-$ guarantees that
both  $E_S$ and $p_S$ are non-negative. Let $\beta_S$ denote the inverse
local temperature  at the shell. Its entropy differential  reads
\begin{equation}
	dS_S = \beta_S \, (dE_S + p_S \, dA_R) \ . \label{dss}
\end{equation}
Use of Eqs. (\ref{m}) and (\ref{ps}) allows us to write the matter
entropy as
\begin{equation}
	dS_S = \beta_S \, ({k_+}^{-1} \, dm_+  -  {k_-}^{-1} \, dm_-) \ .
\end{equation}

How are the three entropies related when the system is in
equilibrium? 
The emergence of equilibrium conditions from general principles within
thermodynamics is the subject of the following section. However,
intuitively 
the  system is in a state of thermal equilibrium provided
(1) $\beta_B = \beta_S \equiv \beta_R$, and 
(2) $\beta_R = \beta_0 \, {k_0}^{-1} \, {k_+}$ at the surface $B_R$. 
The first condition constrains the temperature at the shell 
surface to coincide with the black hole temperature there 
(black hole and shell in mutual thermal  equilibrium), 
whereas the second  guarantees that the total
system is in  thermal equilibrium with its components \cite{To}.
Mechanical equilibrium of the matter shell with the black
hole is  guaranteed by the shell pressure equation (\ref{ps}).
Under these conditions, Eqs. (\ref{ds0}), (\ref{dsb}), and
(\ref{dss}) jointly imply that 
\begin{equation}
	dS_0 = dS_B + dS_S \ .
\end{equation}
The entropy of the composite system is, therefore, additive
with respect  to its constituent subsystems.  Since the entropy is a
function of energy and size (its ``extensive variables" discussed below)
this means that, up to a global constant,  
\begin{equation}
	S_0 (E_0, A_0) = S_B (E_B, A_R) + S_S (E_S, A_R)  \ .
	\label{additivity}
\end{equation}
In the preceding analysis, additivity is a direct consequence of the 
conditions of thermal and mechanical equilibrium.  Observe that it is
independent of the functional  form of the parameters $\beta_S (E_S,
A_R)$ and $\beta_B (E_B, A_R)$. This is as expected, since inverse
temperature appears in the first  law of thermodynamics  as an
integrating factor.  In particular, the foregoing derivation of
additivity does not depend on the special choice of  boundary
conditions or phenomenological matter action  employed in Ref.
\cite{MaYo} or on spherical symmetry \cite{Za1}.
It depends only on the adopted definition of stress-energy tensor
\cite{BrYo1} in terms of quasilocal quantities.

As discussed in the following section, the entropies $S_0 (E_0, A_0)$,
$S_B (E_B, A_R)$, and $S_S (E_S, A_R)$  can be determined from Eqs.
(\ref{ds0}), (\ref{dsb}) and (\ref{dss})  {\it only if}  the
precise forms of all the thermodynamical equations
of state are known. The latter express intensive parameters as  functions
of the appropriate extensive parameters.  For example, it is  well
known that if the thermal equation of state for a black hole is given
by Hawking's  semiclassical expression\cite{Ha}
\begin{equation}
	\beta_B (E_B, A_R) = 8 \pi E_B \, 
	\Bigg( 1 - {{E_B}\over{2R}} \Bigg) 
	\Bigg( 1 - {{E_B}\over{R}} \Bigg) \ , \label{betah}
\end{equation}
then Eq. (\ref{dsb}) yields the Bekenstein-Hawking entropy  \cite{Yo}
\begin{equation}
	S_B (E_B, A_R) = 4 \pi {E_B}^2 
	\Bigg( 1 - {{E_B}\over{2R}} \Bigg)^2  
	= 4 \pi {m_-}^2 \ . \label{sbh}
\end{equation}

These expressions are sufficient to demonstrate that
additivity of entropies for spatially separate subsystems does {\it not}
require the entropy of each constituent system to be a homogeneous
first-order
function of its extensive parameters. This is in contrast to ordinary
thermodynamics \cite{Ca,GrCa}. Recall that a function $f(x_1,...,x_n)$ is
said to be a homogeneous  $m$-th order function of the variables
$(x_1,...,x_n)$ if it satisfies the identity 
\begin{equation}
f(\lambda  x_1,...,\lambda  x_n) = {\lambda}^{m} \, f(x_1,...,x_n) \ ,
\end{equation}
where $\lambda$ is a constant. Upon the
rescaling $E_B \to \lambda E_B$, $A_R \to {\lambda}^2 A_R$ 
($R \to \lambda R $) the entropy $S_B (E_B, A_R)$ in Eq. (\ref{sbh})
behaves as a homogeneous second-order function of $E_B$ and
as a first-order function of $A_R$, namely \cite{Yo,MaYo}
\begin{equation}
	S_B (\lambda  E_B, {\lambda}^2 A_R) = 
	{\lambda}^2 \, S_B (E_B, A_R) \ . \label{sb}
\end{equation}
Equations (\ref{betah}) and (\ref{pb}) also illustrate a central
characteristic of gravitational systems:  the inverse temperature and
pressure are not   homogeneous zeroth-order functions. Under  rescaling
they  behave as \cite{Yo} $\beta_B  (\lambda  E_B, {\lambda}^2 A_R) =
{\lambda} \, \beta_B  (E_B,  A_R) $ 
and
$p_B (\lambda  E_B, {\lambda}^2 A_R) = 
{\lambda}^{-1} \, p_B  (E_B,  A_R) $.

The functional form of $S_S (E_S, A_R)$ and its homogeneous properties 
depend on the explicit form of the matter equations of state. 
These arise from either a phenomenological or a field theoretical 
description of the matter fields involved, and their precise 
form does not concern gravity. Examples of equations of
state for a  self-gravitating matter shell (in the absence of a black
hole)  in thermal equilibrium with itself have been studied
in Ref. \cite{Ma}. Observe that  
Eq. (\ref{p0}) for $p_0$ and Eq. (\ref{pb}) for 
$p_B$ are indeed equations of state while 
Eq. (\ref{ps}) for $p_S$ is not. If the equations of state were known for both 
components, the total entropy $S_0$ could be obtained by 
Eq. ({\ref{additivity}) as a function of the extensive parameters of 
the subsystems. A discussion of this point and of further properties of
this model are delayed  to the following section.

\section{The fundamental problem and the postulates}

The preceding analysis motivates the search for  principles
that are independent of model problems and  that incorporate the
characteristics of gravity  into a logical framework  more general than  
ordinary thermodynamics.  We must start, therefore, by reviewing the basic
assumptions. 

Gravitational thermodynamics describes (effectively) static states of
macroscopic  finite-size self-gravitating systems. As in regular
thermodynamics, it is expected that  very few variables survive the
statistical average  involved in a macroscopic measurement.
What are these macroscopic parameters?
The relationship between thermodynamical and dynamical variables in
gravity has been amply discussed and we refer 
the reader to the literature \cite{BrMaYo,ensembles,BrYoRev,LoWh}. For
our purposes, it is enough to recall the following points.
Firstly,  it has been shown in a wide variety of problems (involving
black holes at finite temperature in interaction with matter fields)
that  the 
thermodynamical energy  coincides  with the quasilocal energy $E$ that
naturally follows from the action of a spatially bounded region. 
If ${^3\!{B}}$ denotes the three-dimensional boundary of the system  
and ${^2\!{B}}$ the two-surface resulting from its intersection with a
spacelike hypersurface  $\Sigma$, the quasilocal
energy is the value of the Hamiltonian that generates unit time
translations  on ${^3\!{B}}$  in the
direction orthogonal to the surface ${^2\!{B}}$
\cite{BrMaYo,BrYo1}. 
We {\it postulate} that this is the appropriate energy variable in
(gravitational)  thermodynamics for {\it all} self-gravitating systems.
Secondly,  the analog of the thermodynamical ``size" of the system 
is the induced two-metric  ${\sigma}_{ab}$ of the two-dimensional
boundary surface  ${^2\!{B}}$ \cite{Yo,WhYo,BrMaYo}.
This property reflects the ``surface character" of gravitational
thermodynamics  and is in part a consequence of the lack of an 
operational definition of ``volume"  in the presence of black holes. 
The size reduces to the surface area $A$ of the two-surface only in the 
case of spherical symmetry \cite{BrMaYo}. For composite systems,
quantities that measure size for internal matter components have  to be
found.   Finally, the remaining macroscopic variables are a
finite number of conserved charges. These may include, for example,
angular momentum \cite{BrMaYo}, suitable combinations of electric 
\cite{RN} or magnetic \cite{Za2} charges, cosmological constant
\cite{BrCrMa}, other types of hair \cite{NuQuSu},  and number of
particles for matter systems \cite{Ma}. (The thermodynamic conjugate
quantities to these parameters are chemical potentials defined at the
boundary of the system \cite{RN,BrMaYo}.)

The existence of these macroscopic parameters motivates the first 
postulate:

\noindent {\bf Postulate I:}\, There exist particular states
(called equilibrium states) of self-gravitating systems that are
completely  characterized macroscopically by the specification of a 
finite set of variables. These variables are the quasilocal energy, size,
and a small number of conserved quantities  (denoted generically by the
symbol $N$).

In ordinary thermodynamics a similar postulate is usually applied to
so-called  ``simple" systems \cite{Ca}. 
These  systems  do not include  gravitational or electromagnetic fields
and are by definition macroscopically homogeneous. The previous
postulate incorporates  strongly  self-gravitating
configurations (with or without matter fields).  As shown
below, these systems may be spatially inhomogeneous.  
Systems describable by these parameters may be termed ``simple" in
gravitational thermodynamics.

Observe that the preceding postulate does not imply that every
gravitational system has equilibrium configurations. Very often a
system does not possess an equilibrium state even though it has 
definite values of energy and other parameters. Rather, the postulate
maintains that equilibrium states, in case they exist, are completely
described by the foregoing finite set of parameters.

The variables $(E, \sigma_{ab}, N)$ that describe a gravitational 
equilibrium state are to be called  {\it extensive} parameters.
Extensive quantities are constructed entirely from the dynamical phase
space variables.
Another essential difference with  usual thermodynamics appears
here: in the latter, the extensive  parameters of a 
composite system equal the sum of their values in each of the subsystems. 
As we illustrate below, this is {\it not}  the case in gravitational 
thermodynamics.

Some  extensive variables  of a self-gravitating system cannot 
be constrained in the conventional thermodynamic sense. 
For example, there exist no walls restrictive with respect to angular
momentum of a stationary black hole system. However,  this is not unusual
or  particular to gravity. For instance, it also occurs in the treatment
of magnetic systems: there exist no walls restrictive  with respect to
magnetic moment. However, one can maintain always the value of magnetic 
moment constant at a boundary surface that delimits the system by a
feedback mechanism that continually adjusts the magnetic field 
\cite{Ca}. The same happens in gravitational thermodynamics:
unconstrainable quantities  can be kept constant at a given boundary
surface by continually monitoring the value of  their respective
conjugate chemical potentials at this surface.  The unavailability of
walls that restrict certain extensive variables is only an idiosyncrasy
that does not  affect the applicability of thermodynamics. As in ordinary
thermodynamics, a boundary that does (does not) allow the flux of heat
can be called diathermal  (adiabatic). Observe that the quasilocal energy
adopted here has a very important property for thermodynamics. 
It is ``macroscopically controllable" in the usual thermodynamic sense: 
it can be ``trapped" by restrictive boundaries and ``manipulated" by
diathermal ones.  
A boundary that does not allow the flow of heat and work can be
called  ``restrictive with respect to  quasilocal energy."   A closed
system is defined as one whose extensive variables  (quasilocal energy,
size, etc.) remain effectively constant at its boundary surface.

It is of course difficult to split a self-gravitating system into
independent ``component" systems in the manner  familiar in
ordinary thermodynamics. Although  one can speak of a ``composite"
system formed by two or more subsystems, the  latter 
interact strongly among themselves.   Clearly,  if a
composite system is closed, the simple systems are not  necessarily so. 
However, internal constraints may exist among the component systems.
These are constraints that  prevent the flow of energy or any other
extensive parameter among subsystems.   For example, in our
model problem internal constraints can  restrict the flow of energy
between the two subsystems (for instance, by fixing a particular
value of $E_B$) or area (by fixing $A_R$). The relaxation of  internal
constraints in an equilibrium composite system will create  processes
that will tend to bring the system to a new equilibrium state.

The central problem of thermodynamics of strongly self-gravitating
systems is, therefore, a reflection of the central problem of ordinary
thermodynamics \cite{Ca},  namely: The determination of the equilibrium
states that will result when internal constraints are removed in a
closed, composite system.

What assumptions are needed in order to solve this problem? Equilibrium
states in gravitational thermodynamics must be characterized by a
simple extremum principle.  As any other thermodynamic system, a
gravitational system  will select, in the absence of constraints, any one
of a number of states, each of which can also be realized in the presence
of a  suitable constraint. Each of these constrained equilibrium 
states corresponds to particular values of the extensive parameters 
of each constituent system and has a definite  entropy. Therefore, the
extremum postulate states that if constraints are lifted, the system
will select the state with the largest entropy. Paraphrasing
Callen:
\vfill\eject

\noindent {\bf Postulate II:} \, There exists a single-valued
function (the  entropy $S$) of the extensive variables of any composite
system, defined for all  equilibrium states, and possessing the following
property. In the absence of internal constraints, the values assumed
by the extensive parameters are those that maximize the entropy over the 
manifold of constrained equilibrium states.

This postulate has to be interpreted carefully. 
Physical equilibrium states correspond to states that extremize the total
entropy  over the manifold of constrained equilibrium states. 
Equilibrium states are, therefore, either maxima, minima, or inflection
points of the entropy. However, in the absence of constraints,   the
extensive parameters of the components in the final equilibrium state
will be those that maximize the entropy.
Postulates I and II not only predict equilibrium states, but
also determine their stability properties.
Equilibrium states corresponding to  maxima of entropy are {\it stable} 
whereas {\it unstable} equilibrium states correspond to extrema other
than maxima.  It is important to emphasize that Postulate II 
makes no reference to nonequilibrium states. Furthermore, it 
implies neither that all equilibrium states of a gravitational system
must have maximum entropy nor that stable states do exist. 
After all, it is common to find systems possessing
equilibrium states that are local minima of entropy. Simple examples
include a nonrelativistic self-gravitating gas in a spherical box or
isothermal stellar systems \cite{BiTr}.

The entropy  as a function of its extensive variables constitutes the 
``fundamental equation" of a self-gravitating system.
The first differentials of the fundamental equation 
define its {\it intensive} variables. 
For systems with a vanishing shift vector, the intensive variables
in the entropy representation are $(\beta, \beta p, \beta \mu)$, where 
$\beta$ denotes inverse temperature, $p$ pressure, and $\mu$ chemical
potentials. Systems possessing a nonvanishing shift require 
functional differentials in the
definitions of their intensive parameters. 
(This happens, for example, in the thermodynamic description  of
stationary geometries \cite{BrMaYo}.)
For general spacetimes the conjugate quantities to the size $\sigma_{ab}$
 are proportional to (minus) the spatial stresses introduced
in Ref. \cite{BrYo1}. The intensive parameters are always functions of the
extensive parameters. The set of functional relationships
expressing intensive in terms of extensive parameters are
the thermodynamical equations of state of a self-gravitating system.  
For example, for static (as opposed to stationary) systems these are
\begin{eqnarray}
	\beta &=& \beta (E, A, N) \ , \nonumber \\
	\beta p &=& \beta p \,(E, A, N) \ ,\nonumber \\
	\beta \mu &=& \beta \mu \,(E, A, N)  \ .\label{eqst}
\end{eqnarray}
As in ordinary thermodynamics, once the fundamental equation 
$S(E, {\sigma_{ab}}, N)$ of a system (or, alternatively, its complete
set of equations of state)  is known, {\it all} its  thermodynamical
information can be obtained from it.

The criteria for global and local stability of equilibrium states in
terms of the entropy function are identical to the ones familiar in
ordinary thermodynamics \cite{Ca,Ma}. In particular, global stability
requires that the entropy hypersurface  $S(E, {\sigma_{ab}}, N)$ lies
everywhere below its family of tangent hyperplanes.

It is possible to express the 
fundamental equation in terms of different sets of independent variables
by performing  Legendre transformations on  the entropy. 
These are the so-called Massieu functions \cite{Massieu,Ca}. 
They play a more fundamental role in gravitational than in
ordinary thermodynamics because they are in a
one-to-one correspondence with actions \cite{ensembles}. 
Their preeminence over ``thermodynamic potentials"  has not been
emphasized sufficiently. (The latter are Legendre transforms of energy
and include the Helmholtz  potential $F$ and Gibbs potential $G$.)
For static systems, Massieu functions include, for example, 
${\cal S}_1 (\beta, A, N) = S - \beta E = - \beta F$,  in which
quasilocal energy has been  replaced by its conjugate entropic intensive
parameter (inverse temperature) as independent variable,  and
${\cal S}_2 (\beta, \beta p, N) =  S - \beta E -  \beta p A = - \beta G $
in which, in addition, the size of the system is replaced by its 
entropic intensive parameter. 
The above equations can be generalized  to stationary geometries
if one recalls that in general it is not possible to choose
all intensive variables constant  at a given choice of two-dimensional
boundary surface  \cite{BrMaYo,BrYo2}. 
The basic extremum postulate is very general and  can be reformulated in
these alternative representations: each Legendre  transform of the
entropy is a maximum for constant values of the transformed (intensive)
parameters \cite{Ca}.

It is important to emphasize that the second postulate incorporates 
not only the so-called first law, {\it but also} the
generalized second law \cite{Be,FrPa,Za3} into a thermodynamic formalism.

Finally, how is the entropy of a composite system
related to the  entropies of the subsystems?
In the previous section we illustrated that entropies are additive
despite strong interaction  between subsystems. On the one hand, 
additivity of entropies does not seem to depend critically on the
particular functional form of the intensive parameters. 
On the other hand, it seems to be  a natural consequence of the
additivity  of actions in a path integral approach to  statistical
thermodynamics \cite{MaYo,Za1}. These reasons motivate us  to assume 
the additivity postulate:

\noindent {\bf Postulate III:}\, The entropy of a composite system
is additive  over the constituent subsystems. Furthermore, it is a
continuous, differentiable, and monotonically increasing function of 
the quasilocal energy $E$.

We emphasize three important points. First, we shall {\it not} assume 
in this  postulate that the entropy of each subsystem is a
homogeneous first-order function of  the extensive parameters. The
postulate is, therefore, more general than the corresponding one in 
regular thermodynamics \cite{Ca,GrCa}. Second, the preceding postulate
implies that all Massieu functions are additive  over component Massieu
functions. As we illustrate below,  this is not the case for
thermodynamic potentials \cite{MaYo}. Third, the monotonic property
implies that the temperature is postulated to be non-negative as in
ordinary thermodynamics.

The logic of the previous section must be contrasted with the present
one: additivity is neither the result of equilibrium conditions among
intensive variables,  nor of the functional form of intensive or
extensive parameters, but a fundamental  assumption. Additivity is valid
{\it even} when the subsystems cannot be considered independent and
strongly interact among themselves. As we show in the following
paragraphs,  equilibrium conditions are indeed a  consequence of the
postulates. 

The preceding postulates are the natural extension of the postulates of
nongravitational  thermodynamics necessary to accommodate the extensive
parameters characteristic of  gravitational systems. Are these postulates
sufficient to solve  the fundamental problem despite strong
interactions among  subsystems? The answer is on the affirmative. 
To illustrate this consider again our model problem in the light of the
logic resulting from the postulates.  We shall determine the  equilibrium
state of the closed, composite system, namely, the  relationships that
must exist among extensive variables of the subsystems for the total
system to  be in thermal and mechanical (and in general, chemical)
equilibrium.  We also shall indicate  how far one can proceed in the
explicit solution  of this problem without assuming particular 
expressions for the equations of state of the subsystems.

By Postulate I, the black hole and matter subsystems are simple systems
characterized by the extensive variables $(E_B, A_B, N_B)$  
and $(E_S, A_S, N_S)$, respectively. The composite system is itself a
simple system and is characterized by the variables $(E_0, A_0, N_0)$.
The size of all systems reduces to the  area
of their  respective surfaces. 
By Postulate II,  the 
fundamental thermodynamical equations in the entropy representation are
the functions $S_0 = S_0(E_0, A_0, N_0)$, $S_B = S_B(E_B, A_B, N_B)$, and
$S_S = S_S (E_S, A_S, N_S)$. 
Postulate III states that $ S_0(E_0, A_0, N_0) = S_B(E_B, A_B, N_B) +  
S_S (E_S, A_S, N_S)$.  
For simplicity and with no loss of generality, we assume  $A_B = A_S
\equiv A_R$ and  the quantities  $N_0$, $N_B$ and $N_S$ to be constant.
The system is considered closed  if its quasilocal energy and area are 
kept effectively constant at the boundary $B_0$,
namely 
\begin{equation}
	E_0 = {\rm const.}; \,\,\, A_0 =  {\rm const.}  \label{closure}
\end{equation}
The fundamental problem is to determine the extensive variables 
$(E_B, E_S, A_R)$ as functions of these constant quantities whenever
equilibrium is attained as a result of  relaxing internal constraints. 
Postulate II establishes that the total entropy of a 
composite system in a state of equilibrium
is an extremum, namely, it does not change as a result of an infinitesimal
virtual transfer of heat or work from one subsystem to the other.
Therefore, in equilibrium
\begin{eqnarray}
	dS_0 = 0 &=& dS_B + dS_S \nonumber \\
		 &=& \beta_B \, (dE_B + p_B \, dA_R) +  
		\beta_S \, (dE_S + p_S \, dA_R)  \ ,\label{ds0a}
\end{eqnarray}
where the second equality  is a consequence of Postulates I and
II. The entropic intensive variables are defined in the conventional
way:  

\noindent $\beta_S (E_S, A_R) \equiv (\partial S_S / 
\partial E_S)_{A_R} $,
$\beta_S  p_S (E_S, A_R) \equiv (\partial S_S / \partial A_R)_{E_S} $; 
$\beta_B (E_B,  A_R) \equiv (\partial S_B / \partial E_B)_{A_R} $, and 
$\beta_B  p_B (E_B,  A_R) \equiv (\partial S_B / \partial A_R)_{E_B} $.

Since the quasilocal energy can be expressed as 
$E_0 = r_0 (1-k_0)$,  the closure equations are
equivalent to the condition 
 $m_+ = {\rm const.}$ Because the  energy $E_B$ refers to the surface
$B_R$ which  coincides with the shell surface, it is easy to see that 
the total quasilocal energy  at $B_R$ is
\begin{eqnarray}
	E_R &\equiv & E_B + E_S \nonumber \\
	    &=& R \, (1 - k_+) \ . \label{er}
\end{eqnarray}
This equation is a consequence of the additivity of quasilocal 
energy discussed in \cite{MaYo,BrYo1}. 
(If the black hole energy $E_B$ is defined at a surface which does not
coincide with  the shell surface,  the energies $E_B$ and $E_S$ are not
simply additive as in Eq. (\ref{er}) \cite{MaYo}.)
The closure conditions and Eq. (\ref{er}) allow the total entropy
(\ref{ds0a}) to  be written as
\begin{equation}
	dS_0 = 0 = (\beta_S - \beta_B)\, dE_S +
           (\beta_B \, p_B + \beta_S \, p_S  - \beta_B \, p_R) \, dA_R 
\label{ds0b} \ ,
\end{equation}
where the pressure $p_R$ is defined as
\begin{equation}
	p_R (E_R, A_R) \equiv {{{E_R} ^2}\over{16 \pi {R}^3}} 
	\Bigg(1 - {{E_R}\over{R}} \Bigg)^{-1} = \,
	{{1}\over{16\pi \, R \, k_+}} (1 - k_+)^2 \ . \label{pr}
\end{equation}
Since the equality in Eq. (\ref{ds0b}) must be satisfied by
independent and arbitrary
variations of $E_S$ and $A_R$, we must necessarily have 
\begin{equation} 
	\beta_S = \beta_B \equiv \beta_R \ , \label{equilb} 
\end{equation}
and
\begin{equation}
	p_S + p_B = p_R \ . \label{equilp}
\end{equation}

The preceding equations are the sought equilibrium
conditions. They state the relationship among intensive variables of the
subsystems  for the composite system to be in thermal and mechanical
equilibrium. As in nongravitational thermodynamics, they yield
a formal solution to the fundamental problem {\it provided}
the equations of state  
\begin{eqnarray}
\beta_B &=& \beta_B (E_B, A_R, N_B) \ , 
\,\, p_B = p_B (E_B, A_R, N_B) \ ; 
\nonumber \\
\beta_S &=& \beta_S (E_S, A_R, N_S) \ , 
\,\, p_S = p_S (E_S, A_R, N_S) 
\end{eqnarray}
for the  subsystems are known. If this is so, Eqs. (\ref{equilb}) and
(\ref{equilp})  are two formal relationships among  $E_B$, $E_S$, $A_R$
and $m_+$  (with $N_S$ and $N_B$ each held fixed). Equations
(\ref{closure}), (\ref{er}),  (\ref{equilb}),  and (\ref{equilp}) are,
therefore, the four desired  equations that determine the four sought
variables  $(E_B, E_S, A_R, m_+)$.

Naturally, the variable $m_+$ in Eqs. (\ref{closure}) and 
(\ref{er}) does not play an important role  in the formalism:
the relationship among the three energies
$E_0$, $E_B$, and $E_S$  can be written explicitly as 
\begin{equation}
	E_0 = r_0 \, \Bigg\{ 1 - \Bigg[ 1 - {{2(E_B + E_S)}\over{r_0}}
	\bigg(1-{{E_B + E_S}\over{2R}} \bigg) \Bigg]^{1/2} \Bigg\}
	 \ . \label{e02}
\end{equation}
For a closed system, Eqs. (\ref{equilb}), (\ref{equilp}),
and (\ref{e02}) (with Eq. (\ref{closure})) provide three desired equations
to determine the three sought  variables $(E_B, E_S, A_R)$.

The fundamental problem is  formally solved by the postulates  {\it
despite} the quasilocal  energy $E_0$ not being the simple sum of the
component  energies $E_B$ and $E_S$ due to binding and self-energy
interactions characteristic of gravitational systems. (It is easy to
see, by using Eq. (\ref{e02}) that all thermodynamic potentials 
obtained from $E_0$  by Legendre transforms are {\it not} the 
simple sum of the  component potentials.) 
This indicates not only that the postulates
form a complete set of assumptions  for a  more general class of
thermodynamic systems than previously considered, but also the
appropriateness of the adopted definitions of extensive parameters.

Consider some further consequences of the postulates. 
Firstly, the mechanical equilibrium condition (\ref{equilp})  represents
the  spatial stress component of the junction conditions at the surface
$B_R$.  (It reduces to Eq. (\ref{ps}) if the pressure equation of state
for the black hole is given by ({\ref{pb})). Secondly, additivity of
entropies and  the equilibrium conditions  (\ref{equilb}) and
(\ref{equilp}) allow the differential of the  total entropy to be written
as  
\begin{eqnarray}
	dS_0 &=&  \beta_0 \, (dE_0 +  p_0 \, dA_0)  \nonumber \\
	     &=&  \beta_R \, (d(E_B + E_S) + (p_B + p_S) \, dA_R) 
		   \nonumber \\
	     &=& \beta_R \, (dE_R + p_R \, dA_R) \ . \label{dst2}
\end{eqnarray}
This expression  confirms that there is no  
``gravitational" entropy  associated to the shell \cite{DaFoPa,MaYo} 
and illustrates that $E_R$ is  the total quasilocal energy and $p_R$ 
the associated pressure associated to the surface $B_R$. 
In turn, Eq. (\ref{dst2}) implies that
\begin{equation}
dS_0 = \beta_0 \, {k_0}^{-1} \, dm_+ = \beta_R \, {k_+}^{-1} \, dm_+ \ .  
\end{equation} 
Therefore, the inverse temperature at the surface
$B_R$ is given in terms of the inverse  temperature $\beta_0$ at the
boundary $B_0$  by \begin{equation}
\beta_0 \, {k_0}^{-1}= \beta_R \, {k_+}^{-1} \ .
\end{equation}  
The (spatially) inhomogeneous character \cite{To} 
of thermodynamic equilibrium (equivalence principle) is  a
consequence of  the postulates of thermodynamics and the definition of
quasilocal stress-energy. Thus, the postulates  do incorporate
equilibrium states in  inhomogeneous systems in contrast to the ordinary
postulates of thermodynamics \cite{Carr}, where a system that is not
homogeneous is not in thermodynamic equilibrium even if its  properties
remain constant in time.

The preceding treatment of a composite self-gravitating system must be 
contrasted with the equivalent one of a composite nongravitational
system presented in Appendix A. Although the systems are physically
different, their similarities and  differences  are readily apparent.  
In particular, the  gravitational equations (\ref{e02}) and
(\ref{closure}) substitute the relations (\ref{etapp}) and (\ref{vtapp})
of flat spacetime thermodynamics.

The formalism provides the methodology to solve the fundamental
problem  for self-gravitating  systems.  In the spirit of
thermodynamics, it yields explicit answers for explicit functional forms
of the fundamental  equations  (or equivalently,  the associated 
equations of state) of each of the subsystems \cite{Ca}. 
These are outside the realm of thermodynamics and are the result
of either phenomenological or statistical mechanical descriptions of the
constituent systems. We reiterate its formal structure: 
for a composite self-gravitating system one must assume the
fundamental equation of the components to be known in principle.  
If the total system is in a constrained
equilibrium state  (characterized by particular values of the extensive
parameters for each constituent  system), the total entropy is obtained
by adding the individual entropies of the components and
is, therefore, a known function of  their extensive parameters.  
The extrema of the total
entropy define the equilibrium states. Stable states correspond to
maxima of entropy.  As an illustration, if explicit
equations of state are known for both black hole and matter in our
model problem,  their entropy values can be found (up to an overall
constant) by integrating Eqs.  (\ref{dsb}) and (\ref{dss}), and
substituting the values of  $(E_B, E_S, A_R)$ at equilibrium. The total
entropy is then  given by  Eq. (\ref{additivity}).

Can this logical framework accommodate 
runaway instabilities (the so-called gravothermal catastrophe) observed
in bounded self-gravitating systems? The answer is on the affirmative. 
This behavior is a consequence of the postulates and the particular form
of the fundamental equations characteristic of gravity. 
Typically, the latter are such that there exist, besides equilibrium
states that locally maximize the entropy, equilibrium states that
locally minimize it. (The existence of these state is well-known in
stellar dynamics \cite{BiTr} and black hole physics \cite{Yo,WhYo}.)
Consider as illustration an  isothermal self-gravitating gas in a closed
spherical container \cite{BiTr}. 
The system might be thought of as formed by two components,
the `core' and the `halo.' The formalism states that  if  the entropies
for the components are known as functions of their extensive parameters, 
the total entropy is $S = S_c + S_h$.  Equilibrium states are
obtained by extremizing this function and are characterized by
particular values of the extensive parameters of the  components.   For
simplicity, consider only the energies (or equivalently,  the density
contrast between components). The entropy functions for the gas 
components are such that there exists an equilibrium state (described by
a particular critical value of the density contrast) that is  a local
entropy minimum over the set of all possible equilibrium states
\cite{BiTr}.  But Postulate II predicts that this state is unstable. The
onset of instability in the gas obeys the postulates: if the system finds
itself in that state and the density contrast between core and halo is
allowed to change, the system will try to reach equilibrium states of
higher entropy.  The system finds it advantageous to transfer energy (or
work) from one region to another, developing more internal
inhomogeneities. Local stability conditions \cite{Ca,Ma} imply that a
negative heat capacity is associated to a local entropy minimum: if the
core gets hotter than the halo, heat flows from the core to the halo and
the core temperature raises. The end result depends on  the form of the
entropy function and on the direction of the fluctuation that started the
instability.  It might be that a local entropy maximum exists in which
the system settles down (as discussed in Ref. \cite{BiTr}, this occurs if
the entropy  is such that, for example, $C_h < |C_c|$). In this case  the
halo temperature rises more than the core's and the system shuts off in a
stable state. A runaway instability  happens if there does not exist a
local  maximum for the system to settle down  (this happens if the
fundamental equation is such that $C_h > |C_c|$). In this case the
temperature difference between  halo and core keeps growing. Whether a
black hole is created or the system runs out of equilibrium before that
occurs, the important point is that, as long as the system can be
described by equilibrium physics, the postulates {\it predict} its
behavior if the fundamental equations of the components are known. The
above argument applies equally to a collection of stars or  other
astronomical systems.

We have studied so far the impact of gravitational extensive variables
in the thermodynamic formalism. 
But the latter is also characterized by the
functional forms of its intensive variables  (equations of state). These
arise from a dynamical theory but their main characteristic is that, in
general, they are no longer homogeneous zeroth-order functions of
extensive parameters  (for instance, the inverse temperature $\beta_B$ in
Eq. (\ref{betah}) is homogeneous first-order in energy and half-order
in area; although the intensive variable  $\beta p$ in the entropy
representation of a static black hole is homogeneous zeroth-order,
this is not the case for other systems \cite{Ma}.)  This implies that the
consequences of the postulates are different than in  ordinary
thermodynamics, particularly the mathematical properties of fundamental 
equations.  Fundamental equations are in general
no longer  homogeneous first-order
functions of their extensive variables. (Alternatively, the
homogeneous properties of fundamental equations in gravitational
thermodynamics imply that intensive variables are no longer 
homogeneous zeroth-order functions.)
This does not affect the formalism itself, but has direct
implications for at least  two formal relationships among thermodynamic
quantities.  Firstly, the so-called Euler relation  is necessarily
different from the one  familiar in ordinary thermodynamics \cite{Yo}. 
An Euler
relation is a consequence of Euler theorem stating  that a homogeneous
function  $f(x_1, ..., x_n)$ of $m$-th order satisfies the equality
\begin{equation}
	m\, f(x_1, ..., x_n) = x_1 \Bigg( {{\partial f}
	\over{\partial x_1}} \Bigg) + \dots + 
	x_n \Bigg( {{\partial f}\over{\partial x_n}}\Bigg)\ . 
\end{equation}
In standard thermodynamics entropy is a homogeneous first-order function 
and  the  Euler relation  is therefore
	$S = \beta E + \beta p V - \beta \mu N$.
In contrast and as an example,  the Euler relation for a static 
charged black hole reads 
\begin{equation}
	S = {{1}\over{2}} \, \beta \, E + \beta \, p \, A  - 
	{{1}\over{2}} \, \beta \, \mu \, N \ . \label{euler}
\end{equation}
Euler relations for hollow self-gravitating thin shells with  
power law thermal equations  of state have been presented in 
Ref. \cite{Ma}.

Secondly, there does not exist a Gibbs-Duhem relation in gravitational
thermodynamics \cite{Ma}. The Gibbs-Duhem relation in ordinary
thermodynamics is a direct consequence of the homogeneous
first-order properties of the fundamental equation and relates the
intensive parameters of a system. It states that the sum of products of
extensive parameters and the differentials of the (conjugate) intensive
parameters vanishes. In the entropy representation it reads
$E d\beta + V d(\beta p) - N d(\beta \mu) = 0$. In contrast, if one
combines the first law with the
Euler relation (\ref{euler}) for a charged black hole one obtains
\begin{equation}
E \, d \beta -  \beta \, dE + 2 A \, d(\beta p) 
- N \, d(\beta \mu) + \beta \, \mu \, dN = 0 \ .
\end{equation}
The reformulation of an Euler relation and the absence of a
Gibbs-Duhem relation set gravity apart from other interactions: even for
magnetic systems the Euler and the Gibbs-Duhem expressions maintain their
usual relationship.

There are no obstacles in applying the preceding formalism 
to any self-gravitating system. These may include not only
nonrelativistic astronomical objects, but also highly relativistic
systems involving general black holes in interaction with matter. 
 The (spherically symmetric) model problem was
used {\it only} as an illustration  because of its simplicity and
transparency.  For  general situations, 
a larger state space is required to  incorporate a larger number of
extensive variables and thermodynamic equilibrium includes equilibrium
under ``interchange of number $N$." It is possible to
employ a condensed notation where the  symbols $X_i$ and  $P_j$  denote
generically all extensive  and intensive parameters, respectively
(excluding energy and inverse temperature),  as in Ref. \cite{Ca}.   In
this way the equilibrium conditions (\ref{equilb}) and  (\ref{equilp})
are  easily  generalized to include all chemical potentials for the
system. Although the treatment of these and other
systems may be technically  difficult, the resolution of the
fundamental problem of thermodynamics obeys the same
logical structure as the one presented here.

We emphasize again that the characteristics of gravitational
thermodynamics are the result of its extensive parameters (which must
include quasilocal quantities like quasilocal energy)  and the particular
homogeneity properties of its intensive variables (equations of state) as
functions of extensive parameters.  This is a more general and rigorous
characterization of the defining properties of gravitational
thermodynamics than, for example, the one presented in Ref. \cite{Fr}
in the context of black holes.
The definitions of extensive and intensive variables as well as the
changes introduced in Callen's postulates  are the minimal changes
necessary to incorporate the  global aspects and nonlinearities
characteristic of the gravitational interaction into a postulatory
formulation of thermodynamics.

To summarize, we have presented the  overall structure and principles 
of a thermodynamic framework that incorporates  self-gravitating
systems.  All the results of standard (gravitational) thermodynamics are 
a consequence of the generalized postulates and the solution of the
fundamental problem and  can be extracted from them by following the
standard procedures described in Refs. \cite{Ca,Gu}. Possible
applications of the formalism include, for example, the description of
quasi-static  reversible and irreversible
processes, alternative representations, phase transitions, etc.  
Finally, a further postulate is usually introduced in standard 
thermodynamics:  the so-called third law. However, the main body of
thermodynamics  does not require this postulate since in the latter there
is no meaning  for the absolute value (and therefore for the zero) of
entropy.  The role and interpretation of this kind of assumption in the
statistical mechanics of gravitation  is the subject of a future
publication \cite{Maip}.

\acknowledgments

It is a pleasure to thank Abhay Ashtekar, Valeri Frolov, Gerald Horwitz,
Werner Israel, Lee Smolin and especially James York for 
stimulating conversations and critical remarks.
Research support was received from the National
Science Foundation Grants No. PHY 93-96246 and No. PHY 95-14240,
and from the Eberly Research Funds of The Pennsylvania State University.

\appendix
\section{}

The simplest closed, composite system in ordinary
thermodynamics consists of two
subsystems separated by a movable diathermal wall \cite{Ca,Carr}. 
The fundamental problem  is to
determine the extensive variables $(E_1, V_1, N_1)$ and $(E_2, V_2, N_2)$ 
of the subsystems whenever  equilibrium is attained. The quantities $E$,
$V$, and $N$ refer to internal  energy, volume, and number of particles,
respectively.  The closure conditions are
\begin{eqnarray}
	E_T &\equiv & E_1 + E_2 = {\rm const.} \ ,  \label{etapp} \\
	V_T &\equiv & V_1 + V_2 = {\rm const.} \ ,  \label{vtapp} \\
	N_T &\equiv & N_1 + N_2 = {\rm const.} \  \label{ntapp}
\end{eqnarray}
The postulates of thermodynamics allow a formal solution of the problem. 
The conditions of equilibrium follow from the equation
\begin{eqnarray}
	dS_T  = 0 &=& dS_1 + dS_2 \nonumber \\
	&=& \beta_1 \, (dE_1 + p_1 \, dV_1 - \mu_1 \, dN_1) +  
            \beta_2 \, (dE_2 + p_2 \, dV_2 - \mu_2 \, dN_2) \nonumber \\
	&=& (\beta_1 - \beta_2)\, dE_1 + (\beta_1 \, p_1 - 
	\beta_2 \,p_2)\, dV_1 - (\beta_1 \mu_1 - \beta_2 \mu_2) 
	\, dN_1  \ .\label{dstapp}
\end{eqnarray}
The first equality is a consequence of the extremum value of entropy
(Postulate II) whereas the second is a consequence of additivity 
of entropies (Postulate III). The last equality follows from the closure
conditions. 
Since the  variations $dE_1$, $dV_1$, and $dN_1$ are independent,
Eq. (\ref{dstapp}) implies that in the entropy representation
\begin{eqnarray}
	\beta_1 &=& \beta_2  \label{bapp} \ , \\
	\beta_1 \, p_1 &=&  \beta_2 \, p_2  \ , \label{papp} \\
	\beta_1 \, \mu_1 &=&  \beta_2 \, \mu_2 \ . \label{muapp}
\end{eqnarray}
If the equations of state 
\begin{eqnarray}
	\beta_a &=& \beta_a (E_a, V_a, N_a) \ , \nonumber \\
	    p_a &=& p_a (E_a, V_a, N_a) \ , \nonumber \\ 
	  \mu_a &=& \mu_a (E_a, V_a, N_a)
\end{eqnarray}
are known for the subsystems (with $a= 1,2$), then Eqs. (\ref{bapp}), 
(\ref{papp}), and (\ref{muapp}) provide three formal
relationships among $(E_1, V_1, N_1)$ and $(E_2, V_2, N_2)$.
Eqs. (\ref{etapp}), (\ref{vtapp}), and (\ref{ntapp}) together with Eqs. 
(\ref{bapp}), (\ref{papp}) and (\ref{muapp}) 
are, therefore, the six desired equations necessary to determine
the six unknown variables.


\end{document}